\documentstyle[aps,epsfig]{revtex}
\begin{document}

\title{Lifetimes of the $\alpha$ decay chains of superheavy element 115 }

\author{D.N. Basu\thanks{E-mail:dnb@veccal.ernet.in}}
\address{Variable  Energy  Cyclotron  Centre,  1/AF Bidhan Nagar,
Kolkata 700 064, India}
\date{\today }
\maketitle
\begin{abstract}

      The lifetimes of the consecutive $\alpha$ decay chains of recently synthesized new element 115, $^{287}115$ and $^{288}115$,  have been calculated theoretically within the WKB approximation using microscopic $\alpha$-nucleus interaction potentials obtained by folding in the densities of the $\alpha$ and the daughter nuclei with a realistic effective interaction. M3Y effective interaction, supplemented by a zero-range pseudo-potential for exchange along with the density dependence, has been used for calculating the nuclear interaction potentials. Spherical charge distributions have been used for calculating the Coulomb interaction potentials. These calculations provide consistent estimates for the lifetimes of the consecutive $\alpha$ decay chains of the superheavy element 115.  

\end{abstract}



      Recently isotopes of the element 115 have been synthesized \cite{r1} as fusion-evaporation residues in the $^{243}Am + ^{48}Ca$ reaction. With a 248 MeV $^{48}Ca$ projectiles three similar decay chains consisting of five consecutive $\alpha$ decays terminated by spontaneous fission have been obsereved. At a higher bombarding energy of 253 MeV with $^{48}Ca$ beam a differrent decay chain of four consecutive $\alpha$ decays terminated by spontaneous fission have been registered. The  $\alpha$ decay energies and half-lives of nine new $\alpha$ decaying nuclei have been measured. The half lives estimated from the Viola-Seaborg formula with Sobiczewski et.al.constants (VSS) \cite{r2} are inconsistent with the experimentally measured half lives which are less than the estimated values for most of the cases. Theoretical calculations in terms of quantum mechanical barrier penetration using microscopically obtained nuclear potentials have been provided. The observed lifetimes of the consecutive $\alpha$ decays originating from the parent isotopes of the synthesized new element 115 are consistent with the theoretical estimates.

      In the present work, the nuclear potentials have been obtained microscopically by double folding the $\alpha$ and daughter nuclei density distributions with a realistic M3Y effective interaction \cite{r3}. This is the ideal procedure of obtaining nuclear interaction energy for the $\alpha$-nucleus interaction. Any liquid drop like properties such as surface energy are basically macroscopic manifestation of microscopic phenomena. A double folding potential obtained using M3Y effective interaction is more appropriate because of its microscopic nature. A potential energy surface is inherently embedded in this description.  The semirealistic explicit density dependence \cite{r4,r5} into the M3Y effective interaction has been employed to incorporate the effects of density dependence. Penetrability of the pre-scission part of the potential barrier provides the $\alpha$ cluster preformation probability \cite{r6}.  

      The microscopic nuclear potentials $V_N(R)$ have been obtained by double folding in the densities of the fragments $\alpha$ and daughter nuclei with the finite range realistic M3Y effective interacion as

\begin{equation}
 V_N(R) = \int \int \rho_1(\vec{r_1}) \rho_2(\vec{r_2}) v[|\vec{r_2} - \vec{r_1} + \vec{R}|] d^3r_1 d^3r_2 
\label{seqn1}
\end{equation}
\noindent
where the density distribution function $\rho_1$ for the $\alpha$ particle has the Gaussian form

\begin{equation}
 \rho_1(r) = 0.4229 exp( - 0.7024 r^2)
\label{seqn2}
\end{equation}                                                                                                                                           \noindent     
whose volume integral is equal to $A_\alpha ( = 4 )$, the mass number of $\alpha$-particle. The density distribution function $\rho_2$ used for the residual cluster, the daughter nucleus, has been chosen to be of the spherically symmetric form given by

\begin{equation}
 \rho_2(r) = \rho_0 / [ 1 + exp( (r-c) / a ) ]
\label{seqn3}
\end{equation}                                                                                                                                           \noindent     
where                        
 
\begin{equation}
 c = r_\rho ( 1 - \pi^2 a^2 / 3 r_\rho^2 ), ~~    r_\rho = 1.13 A_d^{1/3}  ~~   and ~~    a = 0.54 ~ fm
\label{seqn4}
\end{equation}
\noindent
and the value of $\rho_0$ is fixed by equating the volume integral of the density distribution function to the mass number $A_d$ of the residual daughter nucleus. The distance s between any two nucleons belonging to the residual daughter nucleus and the emitted $\alpha$ nucleus is given by 

\begin{equation}
 s = |\vec{r_2} - \vec{r_1} + \vec{R}|
\label{seqn5}
\end{equation}   
\noindent
while the interaction potential between any such two nucleons $v(s)$ appearing in eqn.(1) is given by the M3Y effective interaction. The total interaction energy $E(R)$ between the $\alpha$ nucleus and the residual daughter nucleus is equal to the sum of the nuclear interaction energy, the Coulomb interaction energy and the centrifugal barrier. Thus

\begin{equation}
 E(R) = V_N(R) + V_C(R) + \hbar^2 l(l+1) / (2\mu R^2)
\label{seqn6}
\end{equation}   
\noindent
where $\mu = mA_\alpha A_d/A$  is the reduced mass, $A$ is the mass number of the parent nucleus and m is the nucleon mass measured in the units of $MeV/c^2$. Assuming spherical charge distribution for the residual daughter nucleus and considering the $\alpha$-paticle to be a point charge, the $\alpha$-nucleus Coulomb interaction potential $V_C(R)$ is given by

\begin{eqnarray}
 V_C(R) =&&Z_\alpha Z_d e^2/ R~~~~~~~~~~~~~~~~~~~~~~~~~~~~~~~~~~for~~~~R \geq R_c \nonumber\\
            =&&(Z_\alpha Z_d e^2/ 2R_c).[ 3 - (R/R_c)^2]~~~~~~~~~~for~~~~R\leq R_c 
\label{seqn7}
\end{eqnarray}   
\noindent
where $Z_\alpha$ and $Z_d$ are the atomic numbers of the $\alpha$-particle and the daughter nucleus respectively. The touching radial separation $R_c$ between the $\alpha$-particle and the daughter nucleus is given by $R_c = c_\alpha+c_d$ where $c_\alpha$ and $c_d$ has been obtained using eqn.(4). The energetics allow spontaneous emission of $\alpha$-particles only if the released energy

\begin{equation}
 Q = M - ( M_\alpha + M_d)
\label{seqn8}
\end{equation}
\noindent
is a positive quantity, where $M$, $M_\alpha$ and $M_d$ are the atomic masses of the parent nucleus, the emitted $\alpha$-particle and the residual daughter nucleus, respectively,  expressed in the units of energy.

      In the present work, the half life of the parent nucleus against the split into an $\alpha$ and a daughter nucleus is calculated using the WKB barrier penetration probability. The assault frequecy $\nu$ is obtained from the zero point vibration energy $E_v = (1/2)\hbar\omega = (1/2)h\nu$. The half life $T_{1/2}$ of the parent nucleus $(A, Z)$ against its split into an $\alpha$ $(A_\alpha, Z_\alpha)$ and a daughter $(A_d, Z_d)$  is given by

\begin{equation}
 T_{1/2} = [(h \ln2) / (2 E_v)] [1 + \exp(K)]
\label{seqn9}
\end{equation}
\noindent
where the action integral $K$ within the WKB approximation is given by \cite{r7}

\begin{equation}
 K = (2/\hbar) \int_{R_a}^{R_b} {[2\mu (E(R) - E_v - Q)]}^{1/2} dR
\label{seqn10}
\end{equation}
\noindent
where $R_a$ and $R_b$ are the two turning points of the WKB action integral determined from the equations

\begin{equation}
 E(R_a)  = Q + E_v =  E(R_b)
\label{seqn11}
\end{equation} 
\noindent

      The two turning points of the action integral given by eqn.(10) have been obtained by solving eqns.(11) using 
the microscopic double folding potential given by eqn.(1) along with the Coulomb potential given by eqn.(7) and the centrifugal barrier. Then the WKB action integral between these two turning points has been evaluated numerically using eqn.(1), eqn.(6), eqn.(7), eqn.(8). The zero point vibration energies used in the present calculations are the same as that described in reference \cite{r8} immediately after eqn.(4). The calculations have been done using the density dependent M3Y \cite{r5} effective interaction (DDM3Y) supplemented by a zero-range pseudo potential, representing the single nucleon exchange term \cite{r4}. In DDM3Y the effective nucleon-nucleon interaction $v(s)$ is assumed to be density and energy dependent and therefore becomes functions of density and energy and is generally written as 

\begin{equation}
  v(s,\rho_1,\rho_2,E) = t^{M3Y}(s,E)g(\rho_1,\rho_2,E)
\label{seqn12}
\end{equation}   
\noindent
where $t^{M3Y}$ is the M3Y interaction supplemented by a zero range pseudo-potential 

\begin{equation}
  t^{M3Y} = 7999 \exp( - 4.s) / (4.s) - 2134 \exp( - 2.5s) / (2.5s) + J_{00}(E) \delta(s)
\label{seqn13}
\end{equation}   
\noindent
where the zero-range pseudo-potential representing the single-nucleon exchange term is given by

\begin{equation}
 J_{00}(E) = -276 (1 - 0.005E / A_\alpha ) (MeV.fm^3)
\label{seqn14}
\end{equation}   
\noindent
and the density dependent part has been taken to be \cite{r5}

\begin{equation}
 g(\rho_1, \rho_2, E) = C (1 - \beta(E)\rho_1^{2/3}) (1 - \beta(E)\rho_2^{2/3})
\label{seqn15}
\end{equation}   
\noindent
which takes care of the higher order exchange effects and the Pauli blocking effects.

      The value of the normalization constant C used in the calculations has been kept fixed and equal to unity. All the calculations have been performed with zero angular momentum transfer. The experimentally measured mean values for the released energy Q have been used in the calculations. The energy E appearing in the above equations is the energy measured in the centre of mass of the $\alpha$ - daughter nucleus system and for the $\alpha$ decay process it is equal to the released energy Q. Since the released energies involved in the $\alpha$ decay processes are very small compared to the energies involved in high energy $\alpha$ scattering, $\beta(E)$ has been considered as a constant and independent of energy \cite{r9}. The zero-range pseudo-potential $J_{00}(E)$ is also practically independent of energy for the $\alpha$ decay processes and can be taken as $-276 MeV.fm^3$. After the WKB action integral $K$, given by eqn.(10) has been evaluated, the half lives of the $\alpha$ decays have been calculated using eqn.(9). The density dependent parameter $\beta(E)$, which is supposed to be dependent on energy, has been kept constant and independent of Q and equal to 1.6 \cite{r9}. 

      The results of the present calculations with M3Y effective interaction alone without exchange interaction and the DDM3Y with the pseudopotential have been presented in table-1 below. The quantitative agreement with experimental data is reasonable. Two results are underestimated possibly beacause the centrifugal barrier required for the spin-parity conservation could not be taken into account due to non availability of the spin-parities of the decay chain nuclei. The term $\hbar^2 l(l+1) / (2\mu R^2)$ in eqn.(6) represents the additional centrifugal contribution to the barrier that acts to reduce the tunneling probability if the angular momentum carried by the $\alpha$-particle is non-zero. Hindrance factor which is defined as the ratio of the experimental $T_{1/2}$ to the theoretical $T_{1/2}$ is therefore larger than unity since the decay involving a change in angular momentum can be strongly hindered by the centrifugal barrier. However, as one can see in table-1 that the theoretical  VSS estimates for $T_{1/2}$ largly overestimates, as many as for six cases, showing inconsistencies while the present estimate is inconsistent for only one case where it largely overestimates but still provides much better estimate than that estimated by the VSS systematics. For two cases the lower limits of experimental results for $^{279}111$ and $^{275}109$ provide hindrance factors of about 9 and 2 respectively suggesting that the angular momenta carried by  $\alpha$-particles are non-zero. For rest of the cases it is close to or within the experimental limit. These hindrances are not due to deformations since the deformation energies already get accounted in the Q values and in the calculations experimental Q values have been used. Moreover, the shell effects are implicitly contained in the zero point vibration energy due to its proportionality with the Q value, which is maximum when the daughter nucleus has a magic number of neutrons and protons. Values of the proportionality constants of $E_v$ with $Q$ is the largest for even-even parent and the smallest for the odd-odd one. Other conditions remaining same one may observe that with greater value of $E_v$, lifetime is shortened indicating higher emission rate. Experimental uncertainty in the Q value associated with the $\alpha$ decay from $^{284}113$ can almost account for the overestimation of theoretical lifetime if higher limit for the experimental Q value instead of the mean value be used for calculation.  

\begin{table}
\caption{Comparison between experimental and calculated $\alpha$-decay half-lives using spherical charge distributions for the Coulomb interaction and effective interactions of M3Y and DDM3Y with zero-range pseudo-potential for the nuclear interaction.}
\begin{tabular}{cccccc}
Parent &       & Expt. & Expt.&Theor.$T_{1/2}$ & Theor.$T_{1/2}$      \\ 
 
$Z$&$A$&$T_{1/2}$&$Q(MeV)$&VSS&DDM3Y(M3Y)\\ \hline
&&&&& \\
  115 &287 & $32^{+155}_{-14} ms$ &$10.74\pm0.09$&207 ms  &49(58) ms      \\
&&&&& \\
  113 &283 & $100^{+490}_{-45} ms$ &$10.26\pm0.09$&937 ms &201(235) ms     \\  
&&&&& \\
  111 &279 & $170^{+810}_{-80} ms$ &$10.52\pm0.16$& 45 ms &10(10) ms   \\
&&&&& \\
  109 &275 & $9.7^{+46}_{-4.4} ms$ &$10.48\pm0.09$& 13.7 ms &2.7(2.8) ms  \\  
&&&&& \\
  107 &271 &         (a)                       &$9.07^{(a)}$& 27.1 s  &4.5(5.0) s     \\ 
&&&&& \\
  115 &288 & $87^{+105}_{-30} ms$  &$10.61\pm0.06$& 997 ms &409(495) ms     \\ 
&&&&& \\
  113 &284 & $0.48^{+0.58}_{-0.17} s$ & $10.15\pm0.06$& 4.13 s &1.54(1.84) s    \\ 
&&&&& \\
  111 &280 & $3.6^{+4.3}_{-1.3} s$  &$9.87\pm0.06$& 5.7 s &1.9(2.2) s   \\ 
&&&&& \\
  109 &276 & $0.72^{+0.87}_{-0.25} s$ &$9.85\pm0.06$& 1.44 s&0.45(0.49) s     \\ 
&&&&& \\
  107 &272 & $9.8^{+11.7}_{-3.5} s$ &$9.15\pm0.06$& 33.8 s &9.7(10.8) s     \\ 
\end{tabular}
$(a)$ Taken from reference \cite{r10} since this event was missed in the Dubna experiment \cite{r1}. 
\end{table}

      The half lives for $\alpha$-radioactivity have been analyzed with microscopic nuclear potentials obtained by the double folding pocedure using DDM3Y effective interaction. This procedure of obtaining nuclear interaction potentials is based on profound theoretical basis. The results of the present calculations using DDM3Y supplemented by a pseudo-potential are in good agreement with the experimentally observed data for the half lives of the alpha decay chains of the superheavy element 115. Such calculations reiterates its success of providing reasonable estimates for the lifetimes of nuclear decays by $\alpha$ emissions for the domain of superheavy nuclei.

\end{document}